\documentclass[12pt]{article}

\usepackage{array,dsfont} 
\usepackage{epsfig}
\usepackage{amssymb}
\usepackage{graphics,graphpap}

\renewcommand{\thefootnote}{\fnsymbol{footnote}}

\newcommand{\ds}{\displaystyle}

\begin{document}

\begin{titlepage}
\begin{flushright}\begin{tabular}{l}
IPPP/06/58\\
DCPT/06/116
\end{tabular}
\end{flushright}
\vskip1.5cm
\begin{center}
   {\Large \bf\boldmath Testing QCD
   Sum Rules on the Light-Cone\\[5pt]
   in $D\to(\pi,K)\ell\nu$ Decays}
    \vskip1.3cm {\sc
Patricia Ball\footnote{Patricia.Ball@durham.ac.uk}
}
  \vskip0.5cm
{\em         IPPP, Department of Physics,
University of Durham, Durham DH1 3LE, UK }\\
\vskip2.5cm 


\vskip5cm

{\large\bf Abstract\\[10pt]} \parbox[t]{\textwidth}{
We compare the predictions for the form factors $f_+^{D\to\pi,K}(0)$
from QCD sum rules on the light-cone with recent experimental
results. We find $f_+^{D\to\pi}(0) = 0.63\pm 0.11$, $f_+^{D\to K}(0) =
0.75\pm 0.12$ and $f_+^{D\to\pi}(0)/f_+^{D\to K}(0)= 0.84\pm 0.04$ in
very good agreement with experiment. Although
the uncertainties of the form factors themselves are larger than the
current experimental errors and difficult to reduce, their ratio is 
determined much more accurately and with an accuracy that
matches that of experiment.
}

\vfill


\vspace*{1cm}

\end{center}
\end{titlepage}

\setcounter{footnote}{0}
\renewcommand{\thefootnote}{\arabic{footnote}}

\newpage

\section{Introduction}\label{sec:1}

Exclusive semileptonic decays of $B$ and $D$ mesons are a favoured means of
determining the weak interaction couplings of quarks within the
Standard Model (SM)
because of their relative abundance and, as compared to non-leptonic
decays, simple theoretical treatment.
The latter requires the calculation of form factors by non-perturbative
techniques, the most precise of which, ultimately, will be lattice QCD
simulations. Another, technically much less demanding, but also
less rigorous approach is provided by QCD sum rules on the
light-cone (LCSRs) \cite{LCSR}. While the main motivation for the
calculation of $B\to\pi$ form factors is the determination of
$|V_{ub}|$, see \cite{Vub} for recent analyses, 
$D\to(\pi,K)$ form factors provide
both the possibility to determine $|V_{cd}|$ and $|V_{cs}|$ from
the semileptonic decays $D\to(\pi,K)\ell\nu$ and, due to the
similarity of the calculation, a check of the validity of
$B\to\pi$ form factor calculations. The impressive accumulation of
data on the experimental side, with recent results from
BaBar~\cite{BaBar}, Belle~\cite{Belle},  BES~\cite{BES}, CLEO~\cite{CLEO}
and FOCUS~\cite{FOCUS},  has been
matched by lattice calculations \cite{latt},
whereas the last comprehensive analysis from LCSRs
dates back to 2000 \cite{alexD}. In view of the recent developments in
LCSRs, in particular the  updates on the hadronic
input, that is the light-cone distribution amplitudes (DAs) of $\pi$ and
$K$ mesons of
leading and higher twist, see Ref.~\cite{BBL06}, it is both
timely and instructive to recalculate the corresponding form factors
from LCSRs and confront the results with experimental data. This is
the subject of this letter. 

\section{\boldmath A Light-Cone Sum Rule for $f_+^{D\to\pi,K}(0)$}\label{sec:2}

The key idea of light-cone sum rules is to
consider a correlation function of the weak current and a current with
the quantum numbers of the $D$ meson, sandwiched between the vacuum
and a
$\pi$ or $K$ state. For large (negative) virtualities of these currents, the
correlation function is, in coordinate-space, dominated by distances
close to the light-cone and can be discussed in the framework of
light-cone expansion. In contrast to the short-distance expansion
employed by conventional QCD sum rules \`a la SVZ \cite{SVZ}, where
non-perturbative effects are encoded in vacuum expectation values 
of local operators with
vacuum quantum numbers, the condensates, LCSRs
rely on the factorisation of the underlying correlation function into
genuinely non-perturbative and universal hadron DAs
$\phi$. The DAs are convoluted with process-dependent amplitudes $T_H$,
which are the analogues of the Wilson coefficients in the
short-distance expansion and can be
calculated in perturbation theory. Schematically, one has
\begin{equation}\label{eq:1}
\mbox{correlation function~}\sim \sum_n T_H^{(n)}\otimes \phi^{(n)}.
\end{equation}
The expansion is ordered in terms of contributions of
increasing twist $n$. The corresponding DAs have been studied in
Refs.~\cite{PB98,BBL06}, both for $\pi$ and $K$ mesons and including two-
and three-particle Fock states up to twist 4. The light-cone expansion is
matched to the description of the correlation function in terms of hadrons
by analytic continuation
into the physical regime and the application of a Borel
transformation, which introduces the Borel parameter $M^2$ and
exponentially suppresses contributions from higher-mass states.
In order to extract the contribution
of the $D$ meson, one describes the contribution of other hadron states by
a continuum model, which introduces a second model parameter, 
the continuum threshold $s_0$. The sum rule then yields 
the form factor in question, $f_+$, multiplied by the coupling of the
 $D$ meson to
its interpolating field, i.e.\ the $D$ meson's leptonic decay constant
$f_D$.

LCSRs are available for the $D\to\pi,K$ form
factor $f_+$ to
$O(\alpha_s)$ accuracy for the twist-2 and part of the twist-3
contributions and at
tree-level for higher-twist (3 and 4) contributions
\cite{Bpi,alexD,BZ01,BZ04}. Although these sum rules allow the
prediction of $f_+$ as a function of $q^2$, the momentum transfer to
the leptons, in this letter we only consider the case $q^2=0$. The
reason is that, in contrast to $B$ decays, the range of $q^2$
accessible to LCSR calculations is rather limited in $D$
decays. Following Ref.~\cite{alexD}, one can estimate this range as
$q^2<m_c^2- 2 m_c\chi$, where $\chi$ is a hadronic scale independent
of the flavour of the heavy quark.
In Ref.~\cite{alexD}, $\chi\approx
0.5\,$GeV was chosen, which translates into $q^2<0.6\,{\rm
  GeV}^2$. In Ref.~\cite{BZ04}, we chose $\chi\approx 1\,$GeV for $B$
decays, which translates into $q^2<-0.9\,{\rm GeV}^2$ for $D$
decays. This has to be compared with the kinematic range in $D$
decays: $0\leq q^2\leq (m_D-m_P)^2$, i.e.\ $q^2<3.0\,{\rm GeV}^2$ for
$D\to\pi$ and $q^2<1.9\,{\rm GeV}^2$ for $D\to K$. That is: even in
the optimistic scenario of Ref.~\cite{alexD}, at most 30\% of the
available phase space can be accessed by direct LCSR calculations. The
form factor for larger $q^2$ has then to be extrapolated, using, for
instance, the modified two-pole formula by Becirevic and Kaidalov
\cite{BK}, which is also frequently used in experimental analyses. 
In view of this situation, and the
converging experimental data on the shape in $q^2$, which allows a
direct extraction of $f_+(0)$ from experiment,\footnote{To be more
  precise, it is $|V_{cq} f_+(0)|$ that can be determined from
  experiment. Assuming, however, the SM to be correct, $|V_{cd}|$ and
  $|V_{cs}|$ are related to $\lambda$, the Wolfenstein parameter, and
  known with negligible uncertainty.}
 we decide to focus on
the prediction of $f_+(0)$ only, whose theoretical uncertainty is
smaller than that of the form factor for positive
$q^2$.  We compile the currently available
experimental and theoretical results for $f_+(0)$ in Tab.~\ref{tab1}.
\begin{table}
\addtolength{\arraycolsep}{3pt}
\renewcommand{\arraystretch}{1.2}
$$
\begin{array}{l|lll}
& f_+^{D\to K}(0) &  f_+^{D\to \pi}(0) &  f_+^{D\to
    \pi}(0)/f_+^{D\to K}(0)\\\hline
\mbox{Belle \cite{Belle}} & 0.695\pm 0.023 & 0.624\pm 0.036
     & 0.898\pm 0.045\\
\mbox{BES \cite{BES}} & 0.78\pm0.05 & 0.73\pm 0.15 &
    0.93\pm 0.20\\
\mbox{CLEO \cite{CLEO}} & 0.760\pm 0.012 & 0.670\pm 0.031 & 0.882\pm
    0.050\\
\mbox{FOCUS \cite{FOCUS}} & \mbox{---} & \mbox{---} & 0.85\pm 0.06\\\hline
\mbox{LCSR \cite{alexD}} & 0.91\pm 0.14 & 0.65\pm 0.11 & 0.71 \pm 0.15\\
\mbox{LQCD \cite{latt}} & 0.73\pm 0.08 & 0.64\pm 0.07 & 0.87\pm 0.09\\
\mbox{This Paper} & 0.75\pm 0.12 & 0.63\pm 0.11 & 0.84\pm 0.04
\end{array}
$$
\addtolength{\arraycolsep}{3pt}
\renewcommand{\arraystretch}{1}
\vspace*{-10pt}
\caption[]{\sf Experimental and theoretical values of $f_+^{D\to
    \pi,K}(0)$; LQCD = lattice QCD. All errors have been added in
    quadrature. BaBar has to date only published data on the shape of
    $f_+^{D\to K}(q^2)$, but not the absolute normalisation \cite{BaBar}. 
    The LCSR value for $f_+^{D\to K}(0)$ 
   corresponds to $\overline{m}_s(2\,{\rm GeV}) = (0.10\pm 0.02)\,{\rm GeV}$
    and has been obtained by an interpolation of the results given, in 
    Ref.~\cite{alexD},
    for several values of $m_s$.}\label{tab1}
\end{table}

Although, as mentioned before, the LCSR for $f_+$ has been
investigated in quite a few publications, the actual formula turns
out to be quite complicated and has never been given in a tangible
form. In this letter, we present, for the first time, a compact
formula for $f_+^{D\to P}(0)$, $P=\pi,K$, to tree-level accuracy, 
which makes explicit the suppression factors for contributions of 
higher twist. 
At tree level, one has, to twist-4 accuracy:
\begin{eqnarray}
\lefteqn{\frac{m_D^2 f_D}{m_c}\,e^{-m_P^2/M^2}\,
f_+^{D\to P}(0) = f_P m_c \int_{u_0}^1
  du\, e^{-m_c^2/(u M^2)}\left\{\frac{\phi_{2;P}(u)}{2u}\right.}\nonumber\\
&& + \frac{m_P^2}{m_c (m_{q_1} + m_{q_2})} \,\left[ \frac{1}{2}\,
  \phi^p_{3;P}(u) + \frac{1}{12}\,\left(
  \frac{2}{u}-\frac{d}{du}\right) \phi^\sigma_{3;P}(u)\right.\nonumber\\
&& \hspace*{3.5cm}\left. - \eta_{3P}
  \left(\frac{1}{u} + \frac{d}{du}\right) \int_0^u d\alpha_1
  \int_0^{\bar u} d\alpha_2
  \,\frac{u-\alpha_1}{u\alpha_3^2}\,\Phi_{3;P}(\underline{\alpha})
  \right] \nonumber\\
&&{}+\frac{1}{m_c^2}\left[ -\frac{1}{2} \frac{d}{du} \int_0^u d\alpha_1
  \int_0^{\bar u} d\alpha_2 \frac{1}{\alpha_3}\left( 
  2 \Psi_{4;P}(\underline{\alpha}) -
  \Phi_{4;P}(\underline{\alpha}) + 2 \widetilde\Psi_{4;P}(\underline{\alpha}) -
  \widetilde\Phi_{4;P}(\underline{\alpha})\right)\right.\nonumber\\
&& \hspace*{1.5cm} - \frac{1}{8}\, u
  \,\frac{d^2}{du^2}\, \phi_{4;P}(u) - \frac{1}{2} \frac{d}{du}\left(
  u \int_0^u dv\left\{ \psi_{4;P}(v) - m_P^2 \phi_{2;P}(v)\right\}
    \right)\nonumber\\
&&\hspace*{1.5cm}\left.\left. + \frac{1}{12}\,\frac{d}{du}\left[m_P^2u^2
  \phi_{3;P}^\sigma(u) \right] - \frac{d}{du}\left[m_P^2u^2
  \phi_{2;P}(u)\right] + \frac{1}{8}\, \delta(1-u) 
\frac{d}{du}\,\phi_{4;P}(u)\right]\right\}\label{1}\\
&\equiv &f_P m_c \int_{u_0}^1 du \, e^{-m_c^2/(u M^2)}\left\{ R_1(u) +
  \frac{m_P^2}{m_c (m_{q_1} + m_{q_2})} R_2(u) + \frac{m_P^2}{m_c^2}\,
  R_3(u) + \frac{\delta_P^2}{m_c^2}\, R_4(u)
  \right\}.\nonumber\\[-12pt] \label{2}
\end{eqnarray}
Here $M^2$, the Borel-parameter, and $u_0=m_c^2/s_0$, $s_0$ being the
continuum threshold, are the sum rule specific parameters introduced
above. $m_c$ is the $c$ quark (pole) mass and $m_D$ the $D$ meson
mass. $f_P$ is the light meson's
leptonic decay constant, $m_P$ its mass and $m_{q_i}$ are its valence-quark
masses. $\phi_{n;P}$ and $\Phi_{n;P}$ etc.\ are twist-$n$ two- and
three-particle light-cone DAs of $P$, as
defined in Ref.~\cite{BBL06}; $\eta_{3P}$ is related to the
three-particle twist-3 matrix element $f_{3P}$ and is also defined in
\cite{BBL06}. $u$ is the longitudinal momentum fraction of
the quark in a two-particle Fock state of the $P$ meson, 
whereas $\alpha_{1,2,3}$, with $\sum \alpha_i=1$, are the
momentum fractions of the partons in a three-particle Fock state. The
first neglected term in the light-cone expansion is 
of order $1/m_c^3$. Although
we only write down the tree-level expression for the form factor, radiative
corrections are known for $R_1$ \cite{Bpi} and the two-particle
contributions to $R_2$ \cite{BZ04}, and will be included in the
numerical analysis. All scale-dependent quantities are calculated at
the (infra-red) 
factorisation scale $\mu^2_F = m_D^2-m_c^2$. The term in $\delta(1-u)$
in the last line of (\ref{1}) is a surface term that arises from
Borelisation and continuum subtraction. $R_1$ in
Eq.~(\ref{2}) contains only twist-2  and $R_2$ only twist-3 contributions,
whereas $R_3$ contains mass-corrections to $R_{1,2}$ and $R_4$
genuine twist-4 contributions which are governed by the matrix
element $\delta_P^2$, see Ref.~\cite{BBL06}. The allocation of
$1/m_c^2$ terms in (\ref{1}) to $R_3$ and $R_4$, respectively, is
governed by the explicit factors $m_P^2$ in (\ref{1}) and the
implicit factors $m_P^2$ and $\delta_P^2$ in the DAs as given in 
Ref.~\cite{BBL06}; the reason why
we split the $1/m_c^2$ corrections into two different terms is
because, numerically, $\delta_\pi^2\approx \delta_K^2$ \cite{BBL06}, but
$m_\pi^2 \ll m_K^2$.

It is evident from Eq.~(\ref{2}) that the respective
weight of various contributions is controlled by powers of 
$1/m_c$;\footnote{The $R_i$ themselves are independent of $m_c$.}
the next term in the light-cone expansion contains 
twist-3 and 5 DAs and is 
of order $1/m_c^3$. Nonetheless, (\ref{2}) cannot be
interpreted as $1/m_c$ expansion: for $m_c\to\infty$, the support
of the integrals in $u$ also becomes of ${\cal O}(1/m_c)$, as $1-u_0=
1 - m_c^2/s_0 \sim \omega_0/m_c$, with $\omega_0\approx 1\,$GeV 
a hadronic quantity \cite{Bpi}. 
In this case, the scaling of the various terms in
$m_c$ is controlled by the behaviour of the DAs near the end-point
$u= 1$. For finite $m_c$, however, the sum rules are not sensitive
to the details of the end-point behaviour, see also Ref.~\cite{angi}. 
Numerically, the expansion in terms of $1/m_c$ works very
well for $B$ decays (with $m_c\to m_b$), 
whereas for $D$ decays the chirally enhanced term
multiplying $R_2$ is $\sim 1.5$. We shall come back to that point in
Sec.~\ref{sec:4}.

It is possible to write down a similar sum rule also for $f_+(q^2)$.
The main difference to the case $q^2=0$ is a modification of 
the argument of the
exponential in (\ref{2}), $m_c^2/(uM^2)\to (m_c^2-(1-u)q^2)/(uM^2)$,
and, more importantly, a change of the weight factors with which
$R_{3,4}(u)$ enter:\footnote{Evidently, the $R_i(u)$ also become
dependent on  $q^2$.} 
the terms in $d\phi/du$ are to be multiplied by a factor
$1/(1-q^2/m_c^2)$, and those with $d^2\phi/du^2$ by
$1/(1-q^2/m_c^2)^2$. Hence, for $q^2\to m_c^2$, the power suppression of
higher-twist terms and, consequently, the light-cone expansion breaks
down. This is the reason why the LCSR method is only applicable for
values of $q^2$ which are parametrically smaller than $m_c^2$.

The focus of this letter is on the calculation of both the individual
form factors 
$f_+^{D\to P}(0)$ and the ratio $f_+^{D\to\pi}(0)/f_+^{D\to K}(0)$;
both values have been determined be several experiments, see Tab.~\ref{tab1}.
Whereas the ratio is largely
independent of the precise values of the QCD sum rule parameters
and can
be determined with small uncertainty, very much like
the form factor ratio for $B\to (\rho,K^*)\gamma$ transitions
\cite{BZ06_2}, the value of $f_+^{D\to P}(0)$ itself also depends on
$f_D$. This decay constant has recently been measured with impressive
accuracy by CLEO, $f_D = (222.6\pm 16.7^{+2.3}_{-3.4})\,$MeV
\cite{Artuso}, which is the value we shall use in our calculation. 

Compared with the analysis of Ref.~\cite{alexD}, 
in this letter we implement the 
following improvements:
\begin{itemize}
\item updated values of twist-2 parameters, from both QCD sum rules
  and lattice calculations
  \cite{BZa1,Braunlatt,chris};
\item two-loop evolution evolution of twist-2 parameters \cite{evolution};
\item updated values of light quark masses, leading to a significant 
  reduction of the theoretical uncertainty \cite{lattms,SRms,Leutwyler};
\item inclusion of $O(\alpha_s)$ corrections to the two-particle
  twist-3 contributions \cite{BZ01,BZ04};
\item complete account for SU(3)-breaking in twist-3 and 4 DAs \cite{BBL06}.
\end{itemize}

\section{Hadronic Input}\label{sec:3}

Let us now shortly discuss the hadronic input to Eq.~(\ref{1}).
As for twist-2 DAs, the standard approach is to parametrise them in
terms of a few parameters which are the leading-order terms in
the conformal expansion
\begin{equation}\label{eq:confexp}
\phi_{2;P}(u,\mu^2) = 6 u (1-u) \left( 1 + \sum\limits_{n=1}^\infty
  a^P_{n}(\mu^2) C_{n}^{3/2}(2u-1)\right).
\end{equation}
To leading-logarithmic accuracy the (non-perturbative)
Gegenbauer moments $a_n^P$ renormalize multiplicatively. This feature is
due to the conformal symmetry of massless QCD at one-loop, 
the $a_n^P$ start to mix only at next-to-leading order
\cite{evolution}. Although (\ref{eq:confexp}) is not an expansion in any
obvious small parameter, the contribution of terms with large $n$ to
physical amplitudes is suppressed by the fact that the Gegenbauer
polynomials  oscillate rapidly and hence are ``washed out'' upon
integration over $u$ with a ``smooth'' (i.e.\ not too singular) 
perturbative hard-scattering
kernel. One usually takes into account the terms
with $n=1,2$; the $a_n^P$ are estimated from QCD sum rules, and, since
very recently, lattice simulations. Both are
expected to become less reliable for large-$n$ moments which describe
increasingly non-local characteristics of $\phi_{2;P}$.
As an alternative, one
can build models for $\phi_{2;P}$ based on an assumed fall-off behaviour of
$a_n^P$ for large $n$.  The model of Ball and Talbot (BT)
\cite{angi}, for instance, 
assumes that, at a certain reference
scale, e.g.\ $\mu=1\,$GeV, the even moments $a^P_{2n}$ fall off
as powers of $n$:
\begin{equation}\label{5}
a^P_{2n} \propto \frac{1}{(n + 1)^p}\,.
\end{equation}
BT fix the 
absolute normalisation of the Gegenbauer moments  by the first
inverse moment:
$$\int_0^1 \frac{du}{2u}\,\left\{\phi_{2;P}(u)+\phi_{2;P}(1-u)\right\} 
\equiv 3 \Delta = 3 \left(1 + \sum_{n=1}^\infty a^P_{2n}\right),$$
which can be viewed as a convolution with the singular hard-scattering 
kernel $1/u$ and gives all $a_{2n}^P$ the same (maximum) weight $1$.
The rationale of this model is that the DA is given in terms of only
two parameters, $p$ and $\Delta$, and allows one to estimate the
effect of higher order terms in the conformal expansion of
observables. A similar model can be constructed for odd Gegenbauer
moments. In this letter, we calculate the form factor using both
conformal expansion, truncated after $n=2$, and the BT model, 
normalised by $a^P_1$ and $a^P_2$, respectively, 
and taking into account terms up to $n=9$. It turns out
that the effect of terms with $n>2$ is very small.

\begin{figure}[tb]
$$\epsfsize=0.45\textwidth\epsffile{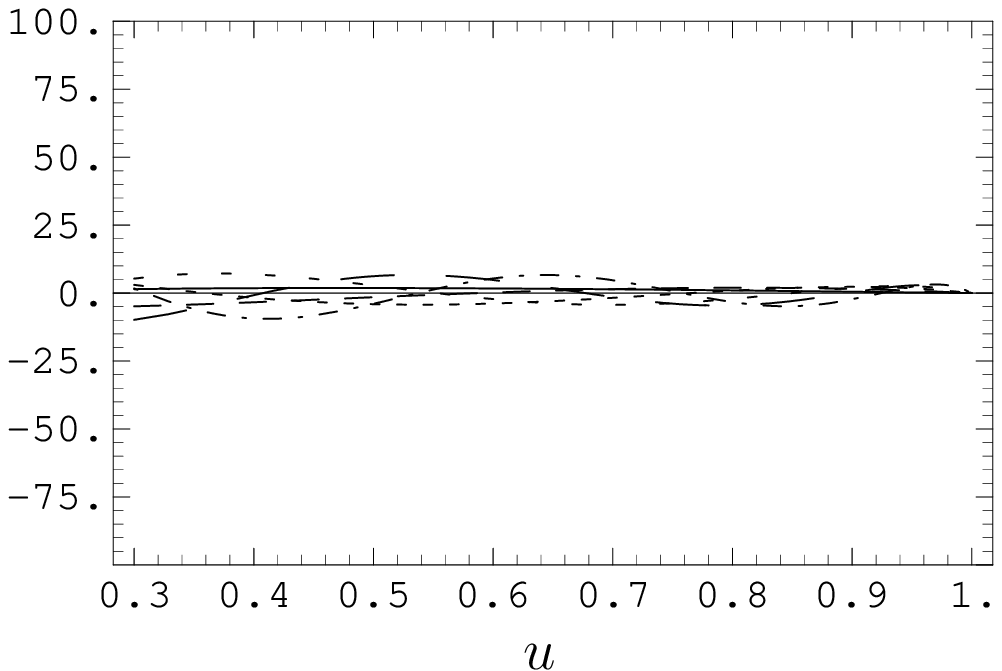}
\qquad \epsfsize=0.45\textwidth\epsffile{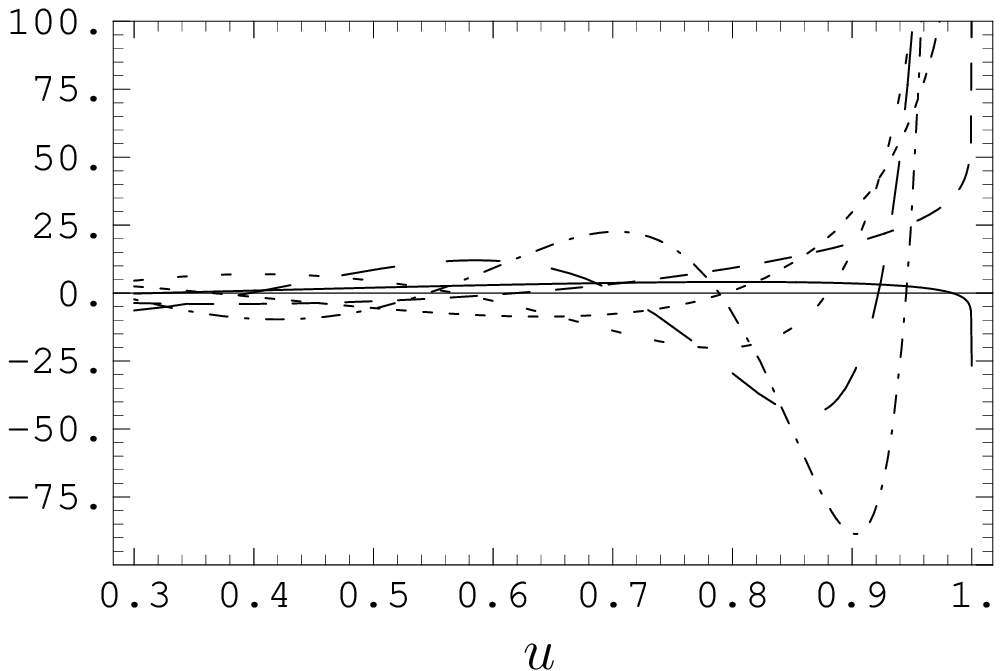}$$
\vspace*{-30pt}
\caption[]{\sf Left panel: contribution of $a_n^P$, $n\leq 5$, to 
$R_1$ as a  function of $u$.
  Right panel: the same for $R_4$ in the renormalon model. The larger
  $n$, the more the contribution of $a_n^P$ 
diverges for $u\to 1$.}\label{fig1}
\end{figure}

As for the numerical values of the Gegenbauer moments,
$a_1^\pi$ vanishes by G-parity and $a_1^K$ has been the subject
of a certain controversy \cite{controversy,KMM}, which has finally
been decided in favour of the value 
\begin{equation}
a_1^K(1\,{\rm GeV}) = 0.06\pm 0.03
\end{equation}
obtained from QCD sum rules \cite{BZa1}. 
Very recently this value has been confirmed from
lattice calculations:
\begin{eqnarray}
a_1^K(1\,{\rm GeV}) & = & 0.057(1)(4)
\quad\mbox{Ref.~\cite{Braunlatt}},\nonumber\\
a_1^K(1\,{\rm GeV}) & = & 0.068(6)
\qquad\mbox{~Ref.~\cite{chris}},
\end{eqnarray}
where we have rescaled the original value given. in
Ref.~\cite{Braunlatt}, 
at the scale $\mu=2\,$GeV by the next-to-leading
order scaling factor
1.26 and the value of Ref.~\cite{chris}, given at $1.6\,$GeV,
by the scaling factor 1.19.
As for $a_2^\pi$, the situation as of spring 2006 is summarised in
Ref.~\cite{BBL06}, with $a_2^\pi(1\,{\rm GeV}) = 0.25\pm 0.15$
averaged over all determinations and $0.28\pm 0.08$ from QCD
sum rules alone. In the meantime, a new lattice calculation has returned
$a_2^\pi(2\,{\rm GeV}) = 0.2\pm 0.1$ \cite{Braunlatt}, which translates into
$a_2^\pi(1\,{\rm GeV}) = 0.3\pm 0.15$. For $a_2^K$, Ref.~\cite{KMM}
quotes $0.27^{+0.37}_{-0.12}$ and Ref.~\cite{BBL06} $0.30\pm 0.15$,
both QCD sum rule results at the scale $1\,$GeV. The first lattice
determination of this quantity is $a_2^K(2\,{\rm GeV}) = 0.18\pm 0.05$
\cite{Braunlatt}, which corresponds to $a_2^K(1\,{\rm GeV}) = 0.26\pm
0.07$. All these results are consistent with each other and indicate
that the values of $a_2^\pi$ and $a_2^K$ are nearly equal. In this letter
we use
\begin{equation}
a_2^\pi(1\,{\rm GeV}) = 0.28\pm 0.08 = a_2^K(1\,{\rm GeV})\,.
\end{equation}
As for twist-3 DAs, we use the expressions and parameters derived in
Ref.~\cite{BBL06}. For twist-4 DAs, i.e.\ the terms entering $R_4$
in (\ref{2}), one can use expressions based
on truncated conformal expansion or the renormalon
model of Ref.~\cite{renormalon}. The advantage of the latter is that
the plethora of independent hadronic twist-4 parameters can all be
expressed in terms of one genuine twist-4 parameter, $\delta_P^2$, and
the twist-2 Gegenbauer moments $a_n^P$. 
One characteristic of the model is that the
end-point behaviour of the DAs for $u\to 0,1$ is more singular than
that of the conformal expansion. While this is a small effect in $B$
decays because of the power-suppression $\sim 1/m_b^2$ of these
contributions, it turns out to be rather problematic in $D$ decays
where the suppression factor is much smaller, and
results in a marked difference in numerics between using the
conformally expanded twist-4 DAs and those based on the renormalon model. In
addition, the conformal expansion converges only badly for the
latter. We illustrate that in Fig.~\ref{fig1}, whose left panel shows
the weight factors with which the Gegenbauer moments $a^P_{n\leq 5}$
enter $R_1$, whereas the right panel shows the corresponding weight
factors for $R_4$. The divergence of the terms in $a_n^P$ for $u\to 1$
is rather striking. The result is a strong dependence of $R_4$, in the
renormalon model, on the order at which $\phi_{2;P}$ is truncated. We
hence decide to drop the renormalon model for the calculation of $D$ decays and
only use the conformally expanded expression for $R_4$.

Other parameters that remain to be specified are the quark masses and
$\alpha_s$. As for the charm quark mass, we use the one-loop pole mass
$m_c = (1.40\pm 0.05)\,$GeV which can be obtained from the value
for $\overline{m}_c^{\overline{\rm MS}}$ found, for instance, from inclusive 
$b\to c \ell\nu$ decays \cite{buchmuller}. For the strange quark mass,
we use $\overline{m}_s(2\,{\rm GeV}) =
(0.10\pm 0.02)\,$GeV, which is in agreement with both lattice
\cite{lattms} and QCD sum rule results \cite{SRms}. As for the
light quark masses, we use the average mass
$\overline{m}_q=(\overline{m}_u+\overline{m}_d)/2$ with
$\overline{m}_s/\overline{m}_q = 24.6\pm 1.2$ 
from chiral perturbation theory \cite{Leutwyler}.
Concerning $\alpha_s$, we use two-loop running down from $\alpha_s(m_Z) =
0.1176\pm 0.002$ \cite{PDG}, which results in $\alpha_s(1\,{\rm
  GeV}) = 0.497\pm 0.005$.

\section{Numerical Results}\label{sec:4}

\begin{figure}[tb]
$$\epsfsize=0.45\textwidth\epsffile{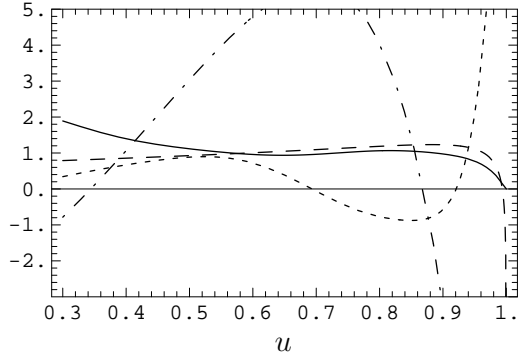}
$$
\vspace*{-30pt}
\caption[]{\sf $R_i$ as function of $u$. Solid line: $R_1$, long dashes:
  $R_2$, short dashes: $R_3$, dash-dotted line: $R_4$.}\label{fig2}
\end{figure}

Equipped with the hadronic input parameters, 
we can now assess the respective size of
the contributions of $R_i$ to the sum rule (\ref{2}). In
Fig.~\ref{fig2}, the $R_i$ are plotted as functions of $u$.
All $R_i$, or at least their integrals over $u$, are of order 1 and hence the
parametric size of their contribution to the LCSR (\ref{2}) 
is indeed set by the weight
factors in (\ref{2}). The central numerical values of these
factors are given in Tab.~\ref{tab2}, together with those for $B$
decay form factors.
\begin{table}
\renewcommand{\arraystretch}{1.2}
\addtolength{\arraycolsep}{3pt}
$$
\begin{array}{l|l|l|l}
& \ds\frac{m_P^2}{m_c(m_{q_1} + m_{q_2})} & \ds\frac{m_P^2}{m_c^2} &
  \ds\frac{\delta_P^2}{m_c^2}\\[8pt]\hline
D\to\pi & 1.44 & 0.01 & 0.08\\
D\to K & 1.41 & 0.13 & 0.09\\\hline
B\to\pi & 0.52 & <0.001 & 0.006\\
B\to K & 0.50 & 0.01 & 0.007
\end{array}
$$
\vspace*{-10pt}
\renewcommand{\arraystretch}{1}
\addtolength{\arraycolsep}{-3pt}
\caption[]{\sf Central values of the weight factors for $R_{2,3,4}$ in
  the LCSR (\ref{2}); the weight factor for $R_1$ is $1$. For
  comparison, the corresponding weights for  $B$
  decay form factors are also shown (based on $m_b = 4.8\,$GeV).}\label{tab2}
\end{table}
Whereas for $B$ decays twist-3 contributions are smaller than
those of twist-2, this is not the case for $D$ decays due to the
chiral enhancement factor $m_P^2/(m_c (m_{q_1} + m_{q_2}))$. This
feature was already noted in Ref.~\cite{alexD} and is a bit
unfortunate from the point of view of using $D\to P$ decays to
  test LCSR predictions for $B\to P$: evidently $f_+^{D\to P}$ is more 
sensitive to the precise value of $1/(m_{q_1} + m_{q_2})$ than
$f_+^{B\to P}$, but the LCSR technique itself is of course completely
independent of that parameter. 
 In addition, $R_2$ is
essentially independent of the Gegenbauer moments $a_n^P$, which enter
$R_2$ only as quark-mass corrections in $m_{q_1}\pm m_{q_2}$, so that
the sensitivity of $f_+^{D\to P}$ to $a_n^P$ is smaller than that of
$f_+^{B\to P}$. Stated differently: a successful calculation of
$f_+^{D\to\pi}(0)$ with a given set of Gegenbauer moments does not
necessarily imply a correct prediction of $f_+^{B\to\pi}$ with the
same moments. The light-cone expansion is also less convergent for
  $D$ than for $B$ decays, so that one may wonder about
  the size of the neglected twist-5 contributions $R_5$, which come with a
  weight factor  $\epsilon^2_{5P}
m_P^2/(m_c^3(m_{q_1}  + m_{q_2}))\sim 1.5\,\epsilon^2_{5P}/m_c^2$ 
with $\epsilon^2_{5P}$ being a twist-5
hadronic matrix element. 

Leaving these reservations aside, at least
for the moment, we proceed to present results for $f_+(0)$.
\begin{figure}[tbp]
$$\epsfsize=0.45\textwidth\epsffile{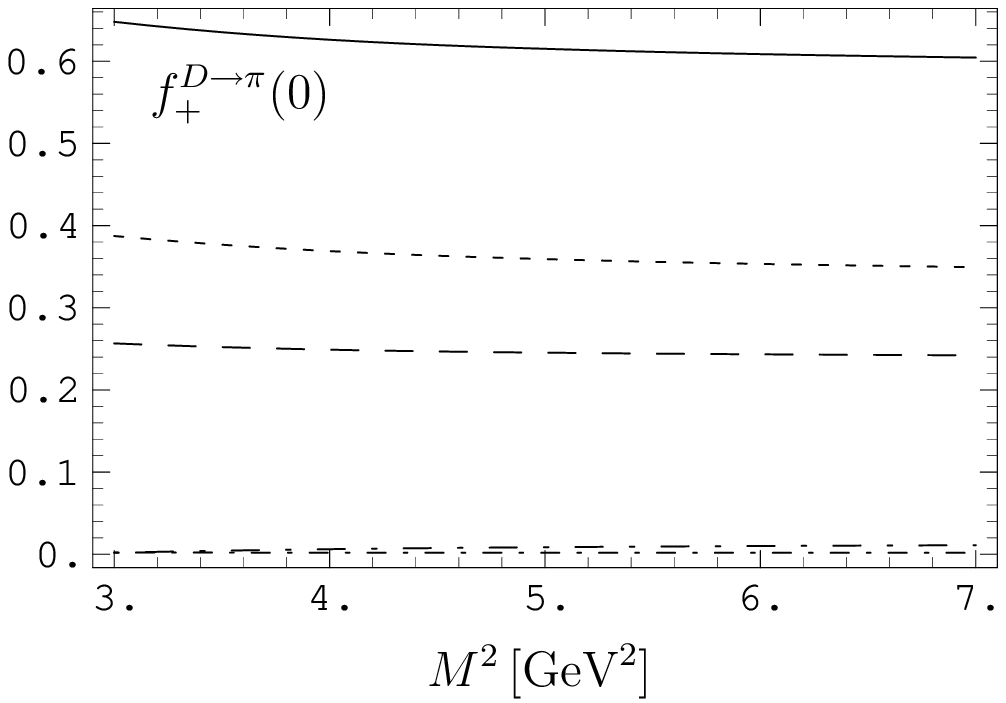}\qquad 
\epsfsize=0.45\textwidth\epsffile{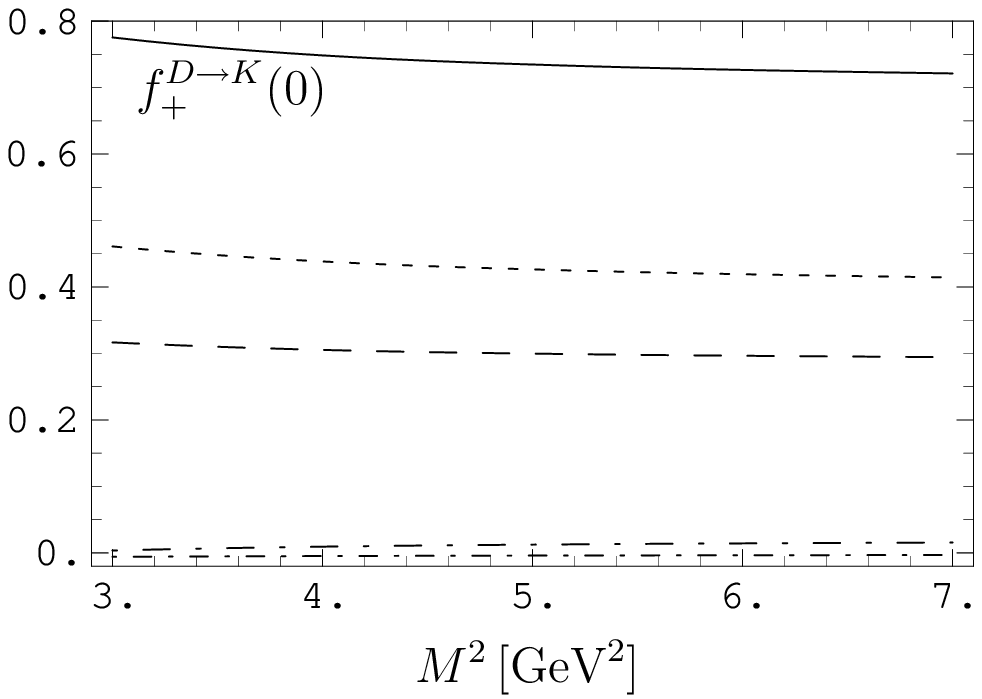}$$
\vspace*{-30pt}
\caption[]{\sf $f_+^{D\to\pi}(0)$ (left panel) and $f_+^{D\to K}(0)$
  (right panel) as functions of the Borel parameter $M^2$ for central
  values of the input parameters. Solid lines: $f_+(0)$, long dashes:
  twist-2 contributions, short dashes: twist-3 contributions,
  dash-dotted lines: twist-4 contributions.}\label{fig3}
$$\epsfsize=0.45\textwidth\epsffile{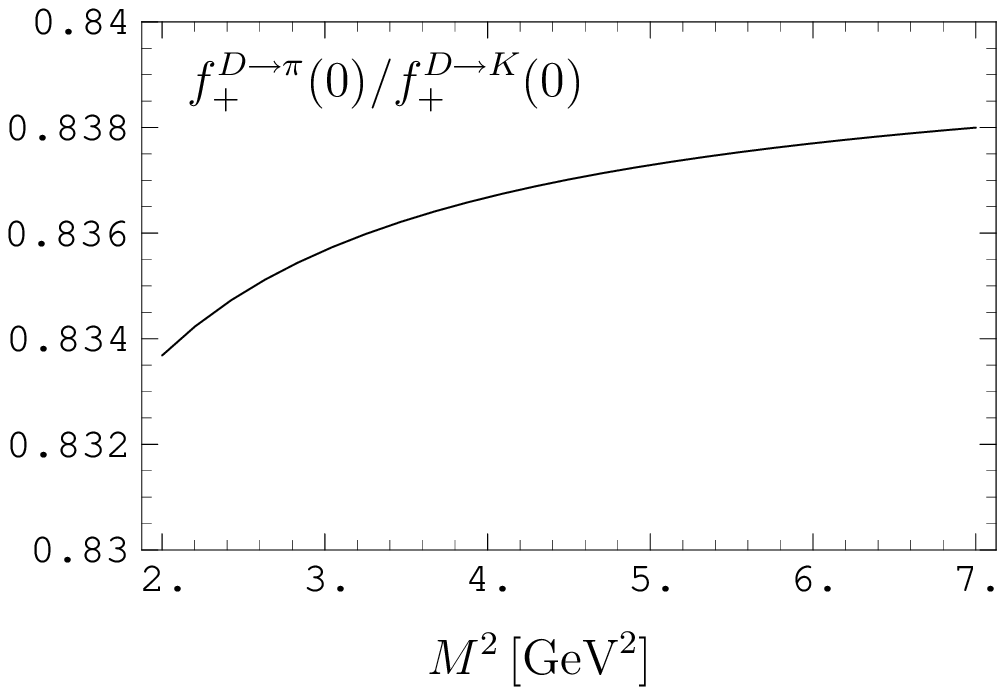}\qquad 
\epsfsize=0.45\textwidth\epsffile{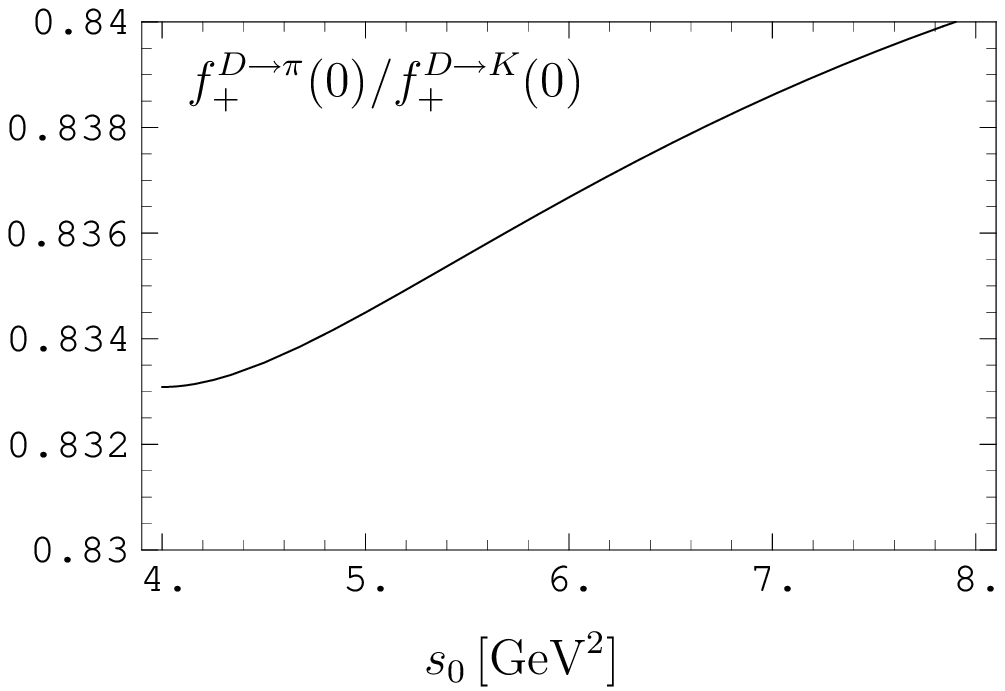}$$
\vspace*{-30pt}
\caption[]{\sf $f_+^{D\to\pi}(0)/f_+^{D\to K}(0)$ as function of
  the Borel parameter $M^2$ (left panel) and
  the continuum threshold $s_0$ (right panel), for central values of
  input parameters.}\label{fig4}
$$\epsfsize=0.45\textwidth\epsffile{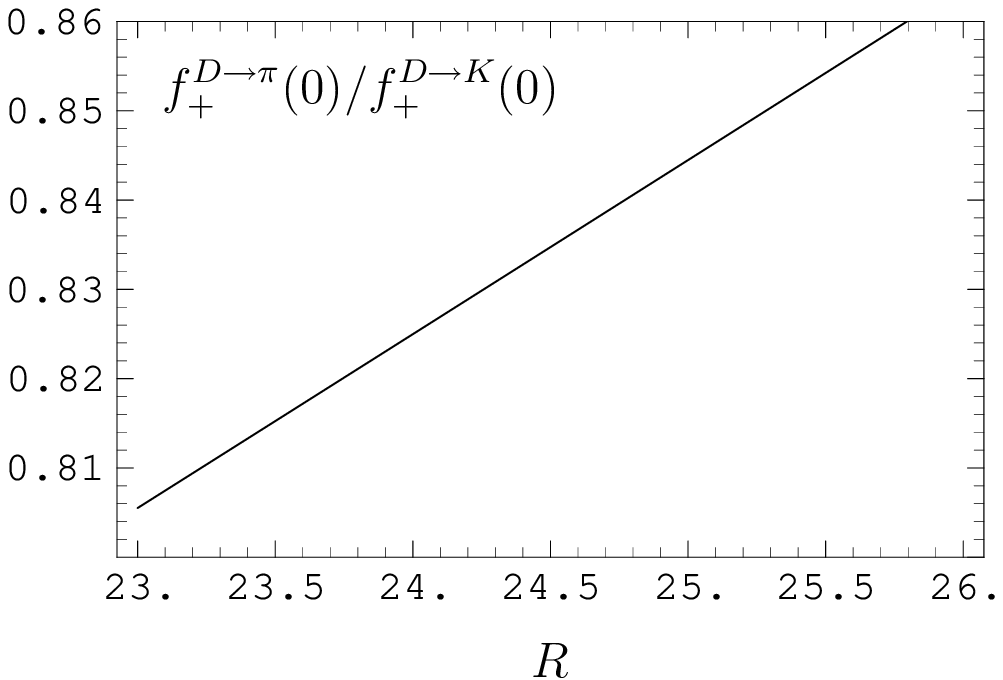}\qquad 
\epsfsize=0.45\textwidth\epsffile{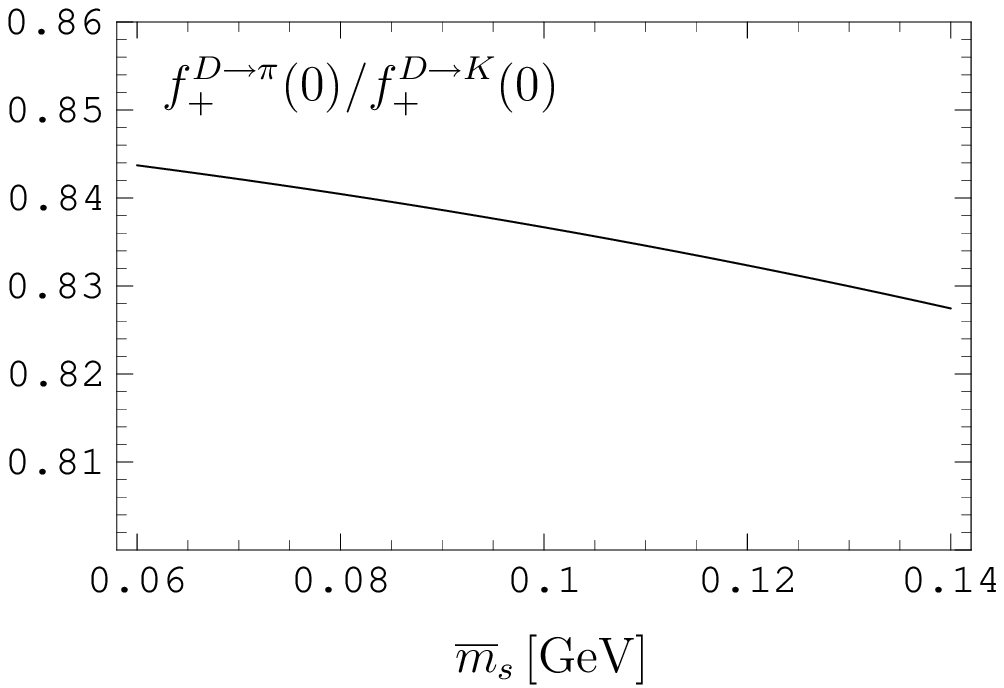}$$
\vspace*{-30pt}
\caption[]{\sf $f_+^{D\to\pi}(0)/f_+^{D\to K}(0)$ as function of
  $R=\overline{m}_s/\overline{m}_q$ (left panel) and
  $\overline{m}_s(2\,{\rm GeV})$ (right panel), for $M^2=4\,{\rm GeV}^2$
  and $s_0=6\,{\rm GeV}^2$.}\label{fig5}
\end{figure}
In Fig.~\ref{fig3} we plot $f_+^{D\to\pi,K}(0)$ as function of the
Borel parameter $M^2$ for $s_0=6\,{\rm GeV}^2$ and central values of
the hadronic input parameters; we also plot the twist-2, 3
and 4 contributions separately. Although the twist-3 contribution is
larger than that of twist-2, due to the chiral enhancement factor,
the hierarchy of higher-twist contributions is preserved and the total
twist-4 contribution is much smaller than that of twist-2 and
3. Fig.~\ref{fig4} shows the ratio $f_+^{D\to\pi}(0)/f_+^{D\to K}(0)$
as function of $M^2$ and $s_0$, respectively. The dependence of the
ratio on these parameters is remarkably small and causes it to vary in
the very small interval $[0.83,0.84]$ only. This is very similar to what we
found in Ref.~\cite{BZ06_2} for the ratio of form factors in $B\to
(\rho,K^*)\gamma$ transitions and due to the fact that the Borel parameter
$M^2$ controls the respective weights of
contributions of different $u$; as these contributions are nearly
equal in numerator and denominator of the ratio of form factors, except for
moderately sized SU(3) breaking, it follows that the resulting dependence on
$M^2$ is very small. 
The parameters to
  which the ratio is most
  sensitive are $R=\overline{m}_s/\overline{m}_q$ and $\overline{m}_s$,
  and we show the corresponding curves in Fig.~\ref{fig5}. Still, the
  dependence of $f_+^{D\to\pi}(0)/f_+^{D\to K}(0)$ on all these
  parameters is very moderate, which allows a very precise prediction
  of this quantity from LCSRs.

In order to obtain final results with a meaningful theoretical
uncertainty, we take $M^2=4\,{\rm GeV}^2$ and $s_0 = 6\,{\rm GeV}^2$ as
our central sum rule parameters and vary both $M^2$ and $s_0$ by $\pm
1{\rm GeV}^2$. We also vary all hadronic input parameters within their
respective ranges as given above or in Ref.~\cite{BBL06}, including
$\alpha_s(m_Z)$ and the factorisation scale $\mu^2$, whose central
value is set to be $m_P^2-m_c^2$. We also include the effect of
switching from the BT model for $\phi_{2;P}$ to a conformal expansion
truncated after the second Gegenbauer moment. Finally, we address the issue of
possible chirally enhanced twist-5 contributions by varying the
twist-4 contributions by a factor 3. When adding all these uncertainties in
quadrature, we obtain the following results:
\begin{eqnarray}
f_+^{D\to\pi}(0) & = & 0.63\pm 0.03 \pm 0.10 = 0.63\pm 0.11\,,\nonumber\\
f_+^{D\to K}(0) & = & 0.75\pm 0.04 \pm 0.11 = 0.75\pm 0.12\,.\label{9}
\end{eqnarray}
Here the first uncertainty comes from the variation of the QCD sum
rule parameters ($M^2$ and $s_0$), the second from the uncertainties
of the hadronic input parameters which are dominated by $f_D$,
$\overline{m}_s$ and $R$. A slight reduction of the total uncertainty
is possible, once more accurate determinations of these parameters will
have become available in the future, but it will be difficult to get below $\pm
0.08$. Our result for $f_+^{D\to\pi}(0)$ nearly coincides with that
obtained in Ref.~\cite{alexD}; this is, however, to a certain extent, 
an accident as quite a few parameters in Ref.~\cite{alexD} were chosen
to have different values, notably $m_c$, $f_D$ and the chiral 
enhancement factor
$m_\pi^2/(2 \overline{m}_q)$, which in Ref.~\cite{alexD} was tied to
the value of the quark condensate. Our value for $f_+^{D\to K}(0)$ is
significantly smaller than that of Ref.~\cite{alexD} as given in
Tab.~\ref{tab1} for the same values of $\overline{m}_s$ we use in this
letter; this is
partially due to the larger $f_D$ we use. The relative errors in
(\ref{9}) are also significantly larger than those quoted, in
Ref.~\cite{BZ04}, for $f_+^{B\to P}(0)$. This is due to the fact
that, for $D$ decays, some parametric uncertainties are larger than
for $B$ decays: the uncertainty due to the light quark masses is three
times larger (see the 2nd column in Tab.~\ref{tab2}); there is a
larger uncertainty due to neglected twist-5 contributions; the
dependence of $f_+$ on $M^2$ and $s_0$ is larger; there is a larger
uncertainty due to $f_D$ for which we use the experimental value
instead of a QCD sum rule.

For the ratio of form factors we find, using the same procedure:
\begin{equation}
\frac{f_+^{D\to\pi}(0)}{f_+^{D\to K}(0)} =  0.84\pm 0.04\,.
\end{equation}
In this ratio, quite a few uncertainties cancel 
so that the total uncertainty is significantly 
smaller than that of
both form factors separately. A reduction of this
uncertainty will be very difficult and requires major progress for
several quantities, including twist-5 contributions.

Our results are in perfect agreement
with the lattice predictions given in Tab.~\ref{tab1}, although our
errors for the form factors are slightly larger. For the ratio of form
factors, our error is by a factor 2 smaller than the lattice
uncertainty quoted in Ref.~\cite{latt}. Comparing with experiment, our
results for the form factors are perfectly consistent with the
experimental results, although the theoretical uncertainty is much
larger than the experimental error quoted by Belle and CLEO. On the
other hand, the
theoretical uncertainty of the ratio of form factors is about the same
size as its experimental counterpart and our predictions agree,
within errors, to
the experimental results.

\section{Summary and Conclusions}\label{sec:5}

The title of this letter is ``Testing QCD
   Sum Rules on the Light-Cone
   in $D\to(\pi,K)\ell\nu$ Decays''. So what is the outcome of this
   test? We have found that the predictions of LCSRs for the form
   factors at zero momentum transfer, $f_+^{D\to\pi}(0)$ and
   $f_+^{D\to K}(0)$, do perfectly agree with both experiment and
   lattice calculations, although the errors are relatively large and
   not expected to be reduced in the near future. The ratio of both form
   factors, on the other hand, can be predicted with much better
   accuracy which matches that of current experimental data and
   surpasses that quoted by the Fermilab/MILC/HPQCD lattice collaboration
   \cite{latt}. Our result agrees within 1.5$\sigma$ with all
   experimental determinations of that ratio, and within 1$\sigma$
   with the experimental average 0.88. 
This indicates that the LCSR method works very well for these form
   factors and with the set of input parameters for $\pi$ and $K$ DAs
   given in Refs.~\cite{BZa1,BBL06,Braunlatt,chris}, and the
   light quark masses $\overline{m}_{q,s}$ obtained from lattice
   calculations \cite{lattms}, QCD sum rules \cite{SRms} and chiral
   perturbation theory \cite{Leutwyler}. This success is certainly
   very encouraging for the LCSR method as such, but unfortunately can
   not be taken as proof that the results for $f_+^{B\to\pi}$ and
   other $B$ decay form factors with the same input parameters will be
   as successful. The main problem area is the larger weight given to
   $a_n^P$ in $B$ decays and the value of $f_B$ which in
   Ref.~\cite{BZ04} was taken from QCD sum rules. Although the $B$
   decay constant has been measured by Belle in early 2006
   \cite{blessedfB}, the experimental uncertainty is yet  too large
   for this measurement to be useful for phenomenology. Nonetheless,
   the overall result is that LCSRs have successfuly passed their
   first serious experimental test in heavy flavour physics and remain
   a serious contender for predicting $B$ decay form factors,
   alongside with and complementary to lattice calculations.

\end{document}